\documentstyle[preprint,aps,eqsecnum]{revtex}
 
\def\beq#1{\begin{equation} \label{#1}}
\def\eeq{\end{equation}}
\def\bra#1{\left\langle #1\right\vert}
\def\ket#1{\left\vert #1\right\rangle}

\begin{document}
{
\tighten
\begin{center}
{\Large\bf CP violation difference in $B^o$ and $B^\pm$ decays
explained \\ 
No tree-penguin interference in 
$B^+ \rightarrow K^+\pi^o$} 
\\
\vrule height 2.5ex depth 0ex width 0pt
\vskip0.8cm
Harry J. Lipkin\,$^{b,c}$\footnote{e-mail: \tt ftlipkin@weizmann.ac.il} \\
\vskip0.8cm
{\it
$^b\;$School of Physics and Astronomy \\
Raymond and Beverly Sackler Faculty of Exact Sciences \\
Tel Aviv University, Tel Aviv, Israel\\
\vbox{\vskip 0.0truecm}
$^c\;$Department of Particle Physics \\
Weizmann Institute of Science, Rehovot 76100, Israel \\
and\\
High Energy Physics Division, Argonne National Laboratory \\
Argonne, IL 60439-4815, USA
} 
\end{center}

\vspace*{0.8cm}
\centerline{\bf Abstract}
\vspace*{4mm}

A new experimental analysis of $B\rightarrow K\pi$ decays provides finite
experimental values for the contributions from interference terms between the
dominant penguin amplitude and the color-favored and  color-suppressed tree
amplitudes. These results can explain the puzzling failure to see CP violation
in $B^\pm\rightarrow K\pi$ decays. Tree-penguin interference contributions are
commonly believed to be the source of the observed direct CP violation in
$B^o\rightarrow K^\pm\pi^\mp$  decays. The data show that the  color-favored
and  color-suppressed  tree contributions interfere destructively  in
$B^\pm\rightarrow K^\pm\pi^o$ decays and nearly cancel. This suprising
cancellation    is not predicted by present theory. There is also no prediction
for any difference  produced by changing the flavor of the spectator quark.
Isospin and Pauli effects that change with spectator quark flavor are examined
and show using group theory and the color-spun SU(6) algebra how they
produce both the near cancellation and the dependence on spectator quark flavor.  
The standard $B\rightarrow K\pi$ analysis which treats tree-penguin
interference only in first order has three parameters overdetermined by four
experimental branching ratios. Previous analyses confirmed the model but with
large errors leaving the values of tree-penguin interference contributions less
that two standard deviations from zero. The new analysis finds interference
contributions well above the errors. 

\vfill\eject

\section {Introduction} 
\subsection {Experiment indicates vanishing of the
tree contribution in $B^+ \rightarrow K^+\pi^o$ decay}

A general theorem from CPT invariance shows\cite{lipCPT} that direct CP
violation can occur only via the interference between two amplitudes which have
different weak phases and different strong phases. This  holds also for all
contributions from new physics beyond the standard model which conserve CPT.
Thus the experimental observation of direct CP violation\cite{Ali} in $B_d \rightarrow
K^+ \pi^-$ and the knowledge that the penguin amplitude is dominant for this
decay require that the decay amplitude must  contain at least one additional
amplitude with both weak and strong phases  different from those of the
penguin. 

This raises the questions what is this ``other amplitude" and what can we learn
about it from experiment.  The failure to observe CP violation in charged
decays\cite{Ali} is still  considered a puzzle\cite{nurosgro}-\cite{ROSGRO}.
One asks the question :``Why should changing the flavor of a spectator quark
which does not participate in the weak decay vertex make a difference?"

After wading through many theoretical papers on the subject I am reminded of my
iconoclastic letter in Physics Today, July 2000, ``Who Ordered Theorists?". The
answers to these questions are already in the experimental data. This paper 
begins by showing how to separate the signal from the noise and find the
answers. The new questions raised by these answers are then addressed using
isospin and permutation symmetry, the Pauli principle and group theory.  

The agreement with experiment\cite{Ali} of the approximate isospin sum
rule\cite{approxlip,approxgr,ketaprimfix} suggests that the dominant ``other
amplitudes" are the color-favored and  color-suppressed tree amplitudes and
that they are sufficiently small to be treated in first order. Second order
contributions are negligible.

There is no new theory at this point. The theoretical implications of the
experiment data are discussed below. There are four experimental branching
ratios available for  $B \rightarrow K\pi$. Therefore three different
independent differences between these branching ratios can be defined which
eliminate the penguin contribution. These can overdetermine the two remaining
free parameters in the theory, the interference contributions between the
penguin amplitude and the color-favored and color-suppressed tree amplitudes.
Unfortunately experimental errors in previous analyses were too large to show
that these parameters differed from zero. 

We choose three differences in a way that minimizes experimental errors and
find that there now are two significant signals in the  data that are well
above the noise of experimental errors and that they still fit an
overdetermination of the two parameters. These indicate the presence of finite
tree-penguin interference contributions that can be the source of the observed
direct CP violation in neutral B-decays.  However the third difference is
consistent with zero well below the noise and below the other two
contributions. The absence of tree-penguin contributions in this difference is
completely unpredicted, provides a new challenge to theorists and can explain
the failure to observe direct CP violation in charged B decays.   

A difference between the tree-penguin interference contributions to charged and
neutral decays is already shown in the experimental analysis of the difference
rule \cite{ketaprimfix} eq. (1.13) of ref. \cite{ketaprimfix}. 

\beq{expeqapp}
2\cdot (12.1\pm 0.8) - (24.1 \pm 1.3) = 0.1 \pm 2.1 \approx  (18.2 \pm 0.8) - 
2\cdot (11.5 \pm 1.0) = - 4.8 \pm 2.2 
\eeq 

The charged decay contribution on the left hand side is zero, with an
unfortunately large experimental error. The neutral decay contributions on the 
right hand side are finite and a bit more than two standard deviations from
zero. But this difference was still not convincing.

The new data analysis sharpens this difference by isolating the color-favored
and color-suppressed  contributions. Both tree-penguin interference 
contributions are now shown to be appreciable and well above the background of
experimental errors. 
The reason that changing the flavor of the spectator quark makes a difference
is that color-flavored and color-suppressed contributions are incoherent when
the  flavor of the spectator quark is different from the flavor of the $u$
quark created in the $b \rightarrow u$ transition. This is the case in the
neutral $B$ decays where the spectator is a $d$ quark. But in $B^+ \rightarrow
K^+\pi^o$, the spectator quark is also a  $u$ quark and the  color-flavored and
color-suppressed  contributions are coherent and interfere. The data now tell
us that these  contributions are finite and well above the experimental errors
but that they are equal and opposite and cancel in $B^+ \rightarrow K^+\pi^o$. 
Thus tree-penguin interference can explain both the presence
of CP violation in neutral decays and its absence charged decays.

New questions now arise because no present theory suggests this cancellation. 

\subsection {Search for possible explanations for the cancellation}

We now look for reasons for this surprising cancellation and for the
difference produced by the flavor of the spectator quark. In particular we
wonder about some symmetry which produces a selection rule that cancels the
tree contribution to $B^+ \rightarrow K^+\pi^o$. 

We first note that changing the flavor of the spectator quark produces a large
isospin difference which may lead to some understanding and might lead to an
isospin selection rule. 

The tree diagram for $B^+ \rightarrow K^+\pi^o$ has a four-body  $u\bar s u\bar
u$ state containing a $u$ spectator quark and the $\bar u u \bar s $ produced
by the $\bar b$ antiquark weak decay.  Combining the two $u$ quarks in an $I=1$
isospin state with the $(I=1/2)$ $\bar u$ antiquark gives a unique four-body
``tetraquark" state with unique isospin couplings.  It is a definite mixture 
of two eigenstates of the total four-body isospin with $I=1/2$ and $I=3/2$ with
unique relative magnitudes and phase. In the ``fall-apart" model\cite{barnes} 
this isospin constraint leads to a selection rule forbidding the tree
contribution to the $B^+ \rightarrow K^+\pi^o$ decay.  

The isospin of the two quarks $(u,d)$ in the corresponding tree diagram for 
the neutral $B$ decays is not unique, it is a combination of $I=0$ and $I=1$.
Thus there is no  isospin constraint here and no selection rule forbidding the
tree contribution. Changing the flavor of the spectator quark makes a crucial
difference in the isospin analysis.

How can any selection rule arise purely from isospin apparently independent of
the nature of the color-favored and color-suppressed transtions? The $u\bar s
u\bar u$ state contains two  identical $u$ quarks which must satisfy the Pauli
principle.  The definition of color-favored and color-suppressed tree diagrams
treats the $u$ quark produced in the weak vertex and the spectator $u$ quark as
distinguishable particles, ignoring the Pauli principle. The Pauli interchange
in the state $u\bar s u\bar u$ transforms between color-favored and  
color-suppressed tree diagrams, implying a symmetry.

The Pauli effect is investigated using the SU(6) color-spin
algebra, denoted here by $SU(6)_{cs}$. The $u\bar s
u\bar u$ ``tetraquark" state which fragments into an s-state of two pseudoscalar
mesons is shown to be in the singlet state of $SU(6)_{cs}$ and restricted by
the Pauli principle to be in a unique state created by the
product $15\otimes\bar {15} $ and in a flavor SU(3) state classified in the 27
dimensional representation with the eigenvalue $V=2$ of the $V$ spin subgroup of
SU(3). 
     
The same tree diagram for $B^o$ decays with a $d$ spectator quark creates a 
$u\bar s d\bar u$ tetraquark state with no Pauli restrictions. The states 
classified in the product $21\otimes\bar {21} $ of $SU(6)_{cs}$ and in a
flavor SU(3) octet which are Pauli-forbidden for $B^\pm$ decays are favored
here.  This can produce a drastic difference between the $B^o$ and $B^\pm$
decays. We also note that it is only the two-pseudoscalar final state that is 
in the singlet state of $SU(6)_{cs}$. Other final states including vector mesons
are classified in other representations of $SU(6)_{cs}$. Thus our treatment does
not apply to these other final states.

\section{Experimental analysis of $B \rightarrow K\pi$}

\subsection {A new anaysis of the data pinpointing tree-penguin interference}
We first show explicitly that the present $B \rightarrow K\pi$ data do
suggest a possible symmetry or selection rule.

We begin with a conventional analysis expressing the four  $B \rightarrow K\pi$
amplitudes in terms of the three amplitudes $P$, $T$ and $S$
denoting respectively the penguin,  color favored tree and color suppressed
tree amplitudes while neglecting other contributions at this
stage\cite{approxlip,approxgr,ketaprimfix}.

\begin{equation}
\begin{array}{ccl}
\displaystyle
A[K^o\pi^+]=P; ~ ~ ~ A[K^+\pi^-]= T +P
\hfill\\
\\
\displaystyle
A[K^o\pi^o]={{1}\over{\sqrt{2}}} [S - P]; ~ ~ ~ 
A[K^+\pi^o]={{1}\over{\sqrt{2}}} [T + S + P] 
\end{array}
\end{equation}

We first note that a selection rule that eliminates the tree contribution to 
$B^+ \rightarrow K^+\pi^o$  predicts that both $B^+$ decays are pure penguin
decays to the $I=1/2$ $K\pi$ state. Experiment\cite{HFAG} shows  agreement 
with this prediction to  between one and two standard deviations.

\beq{bplusselec}
2B(B^+ \rightarrow K^+ \pi^o) = 25.66\pm 1.18 \approx  
B(B^+ \rightarrow K^o \pi^+ ) = 23.40 \pm 1.06
\eeq
where $B$ denotes the branching ratio in units of $10^-6$

We now get a more sensitive tests by using all the $B \rightarrow K\pi$ data.

The agreement with experiment\cite{Ali} of the approximate isospin sum
rule\cite{approxlip,approxgr,ketaprimfix}  tells us that the two tree
amplitudes $T$ and $S$ are sufficiently smaller than the dominant penguin
amplitude $P$ and can be treated only to first order.

We now improve on the previous analysis\cite{ketaprimfix} which converted the
sum rule to a ``difference rule" and obtained eq. (\ref{expeqapp}). We  use new
data and define new differences which optimize the signal to noise ratio.
Noting that the branching ratio $B(B^o \rightarrow K^+ \pi^-) $ has the
smallest experimental error, we define three independent differences which
vanish for a pure penguin transition and are chosen to have the smallest
experimental errors.

\begin{equation}
\begin{array}{ccl}
\displaystyle
\Delta (K^o\pi^+) \equiv |A[K^o\pi^+]|^2 - |A[K^+\pi^-]|^2 \approx - 2 
\vec P \cdot \vec T
\hfill\\
\\
\displaystyle
\Delta (K^+\pi^o) \equiv 2 |A[K^+\pi^o]|^2 - |A[K^+\pi^-]|^2 \approx 2 
\vec P \cdot \vec S
\hfill\\
\\
\displaystyle
\Delta (K^o\pi^o) \equiv 2 |A[K^o\pi^o]|^2 - |A[K^+\pi^-]|^2 \approx - 2 
\vec P \cdot (\vec T+\vec S) 
\end{array}
\end{equation}
where the appoximate equalities hold to first order in the $T$ and $S$ 
amplitudes.
The isospin sum 
rule\cite{approxlip,approxgr} is easily expressed in terms of these differences,

\beq{eqapprev}
\Delta (K^o\pi^o) + \Delta (K^+\pi^o) - \Delta (K^o\pi^+) 
  \approx 0
\eeq

Since each of the three terms in eq. (\ref{eqapprev}) vanish for a pure penguin
transition, the sum rule is trivially satisfied in this case. We shall see that
we can do better than the previous analysis\cite{ketaprimfix}  that only showed
that the sum rule was still only trivially satisfied with real data and that
all terms proportional to tree-penguin interference were still statistically
consistent with zero.  

We now check whether these individual differences are sufficiently different 
from zero with available experimental branching ratio data corrected for the
lifetime ratio\cite{HFAG} 

\begin{equation}
\begin{array}{ccl}
\displaystyle
{{\tau^o}\over{\tau^+}}\cdot B(B^+ \rightarrow K^o \pi^+) - 
B(B^o \rightarrow K^+ \pi^-)  =
2.04 \pm 1.17 \propto -\vec P \cdot \vec T
 \hfill\\
\\
\displaystyle
{{\tau^o}\over{\tau^+}}\cdot 2B(B^+ \rightarrow K^+ \pi^o) - 
B(B^o \rightarrow K^+ \pi^- ) =  
4.15 \pm 1.27 
\propto\vec P \cdot \vec S
\hfill\\
\\
\displaystyle
2B(B^o\rightarrow K^o \pi^o) - B(B^o \rightarrow K^+ \pi^-) =
-.05 \pm 1.41
\propto - \vec P \cdot (\vec T + \vec S)
\end{array}
\end{equation}

The data are now sufficiently precise to show  that
the interference terms between the dominant penguin amplitude
and the color-favored and color-suppressed amplitudes are both individually
finite and one is well above the experimental errors. The sum
rule is satisfied and is now nontrivial. But the term $\Delta (K^o\pi^o)$ which
is proportional to  $\vec P \cdot (\vec T+\vec S)$ is equal to zero and now
well within the experimental errors. This suggests some symmetry or selection
rule.

Thus tree-penguin interference can explain the observed CP violation in
charged B-decays and its absence in neutral decays. 

But there has been no theoretical prediction for this surprising cancellation.

\subsection {A recent restatement of the old sum rule which misses some physics}

The approximate isospin sum
rule\cite{approxlip,approxgr,ketaprimfix} has recently been 
rearranged\cite{Giorgi,Buras} without noting the implications of the present
paper.

\beq{giorgi}
R_n \equiv {{\Gamma(K^+ \pi^-)}\over{2\Gamma(K^o\pi^o)}} = 0.99 \pm 0.07
~ = ~ R_c \equiv {{2\Gamma(K^+ \pi^o)}\over{\Gamma(K^o\pi^+)}} = 1.11 \pm 0.07
\eeq

This restatement of the original agreement with the sum rule\cite{Ali} concludes
that this agrees with the standard model and that the ``$K\pi$ 
puzzle" is no more.
This particular rearrangement and its interpretation misses two crucial points.
\begin{enumerate}
\item The two sides of the equation not only agree; they are both equal to unity
which is the value for the case of a pure penguin transition. The conclusion at
this point is that
any tree-interference is down in the noise of this experiment.
\item The difference between each side and unity is proportional to the
interference between the penguin and the sum of
the two tree contributions,
$\vec P \cdot (\vec T+\vec S)$ 
which we have seen
is equal to zero well within the experimental errors. 
\end{enumerate}

\begin{equation}
\begin{array}{ccl}
\displaystyle
R_n \equiv {{\Gamma(K^+ \pi^-)}\over{2\Gamma(K^o\pi^o)}} = 
{{|\vec P + \vec T|^2}\over {|\vec P - \vec S|^2}} \approx 
1 + 2\cdot {{\vec P \cdot (\vec T + \vec S)}\over {P^2}} 
\hfill\\
\\
\displaystyle
R_c \equiv {{2\Gamma(K^+ \pi^o)}\over{\Gamma(K^o\pi^+)}} =
{{|\vec P + \vec T + \vec S|^2}\over {|\vec P |^2}} \approx 
1 + 2\cdot {{\vec P \cdot (\vec T + \vec S)}\over {P^2}} 
\end{array}
\end{equation}

Our analysis shows that no deviations from unity appear on either side of 
the relation (\ref{giorgi}) because the two tree contributions cancel. That the
two contributions are both indeed finite and interesting is missed in this way
of presenting the sum rule. 

\section{Some isospin arguments searching for a selection rule} 

To try to understand a possible theoretical basis for this accidental
cancellation of the color-favored and color-suppressed contributions to 
$B^+ \rightarrow K^+ \pi^o$ we examine an isospin analysis of the initial and 
final states. 

In the tree diagram the $\bar b \rightarrow \bar u u \bar s $ decay 
together with a $u$ spectator quark produce a
$u\bar s u\bar u$  state.   
This ``tetraquark" state contains two $u$ quarks with
isospin $I=1,I_z=1$. The $\bar u$ antiquark is in a well  defined isospin state
with $I=1/2,I_z=-(1/2)$ and the strange antiquark has  isospin zero. The total
four-body state is thus a state with well defined isospin. It is a definite
linear combination of states with $I=1/2$ and $I=3/2$ with relative magnitudes
and phases determine by isospin Clebsch-Gordan coeffiecients for coupling two
states with $I=1$ and $I=1/2$ to  $I=1/2$ and $I=3/2$.

The $K\pi$ states are also linear combinations of states with isospin (1/2) and
isospin (3/2) with relative amplitudes and phases determined completely by the
requirement that the pion has isospin one and the kaon has isospin (1/2).  The
relevant Clebsch-Gordan coefficients for the $K^+\pi^o$ state  are seen to be
just those to make this linear combination exactly orthogonal to the
combination in the $u\bar s u\bar u$ state produced in the tree diagram 
by the $\bar b$ decay . 
The overlap between the $K^+\pi^o$ state and the initial state thus vanishes. 
In the ``fall-apart"\cite{barnes} model commonly used in tetraquark decays
this vanishing overlap indicates that  the
$B^+ \rightarrow K^+\pi^o$ transition is forbidden. 

We now investigate this explicitly by expanding the initial and final states in
isospin eigenstates,

\beq{initial}
\ket{i;u\bar s u\bar u} \propto \ket{{1\over 2},{1\over 2}}
\bra{1{1\over 2}1(-{1\over 2})}
\ket{1{1\over 2}{1\over 2}{1\over 2}}  + 
\ket{{3\over 2},{1\over 2}}
\bra{1{1\over 2}1(-{1\over 2})}
\ket{1{1\over 2}{3\over 2}{1\over 2}}  
\eeq

\beq{finalkp}
\ket{f;\pi^oK^+} \propto \ket{{1\over 2},{1\over 2}}
\bra{1{1\over 2}0({1\over 2})}
\ket{1{1\over 2}{1\over 2}{1\over 2}}  + 
\ket{{3\over 2},{1\over 2}}
\bra{1{1\over 2}0({1\over 2})}
\ket{1{1\over 2}{3\over 2}{1\over 2})}  
\eeq
where $\bra{j_1j_2m_1m_2}\ket{j_1j_2JM}$ denotes a Clebsch-Gordan coefficient. 
 From the orthogonality relation for Clebsch-Gordan coefficients
\beq{ortho}
\langle f;\pi^oK^+ \ket{i;u\bar s u\bar u} =0.
\eeq

There is therefore no overlap between the initial ``tetraquark" state 
$\ket{i;u\bar s u\bar u}$ produced by the weak interaction and the $K^+\pi^o$  
final state. 

The transition is therefore forbidden if the tetraquark state that fragments
into a kaon and a pion is the same as the original tetraquark state created in
the tree diagram by the $\bar b$ antiquark decay; i.e. if the relative amplitude and
phase between the $I=1/2$ and $I=3/2$ components are preserved between the
creation and fragmentation of the tetraquark. This is true in the simple 
``fall-apart"\cite{barnes} decay mode in common tetraquark models. The
experimental data seem to indicate that the transition is indeed forbidden here.

The conventional analysis which does not include Pauli antisymmetrization has
the amplitude for this transition given by the sum of  independent color
favored and color suppressed amplitudes. The experimentally observed vanishing
of this transition suggests that these amplitudes must cancel for the decay
$B^+ \rightarrow K^+ \pi^o$. It will be interesting to check whether this
cancellation is required when Pauli antisymmetrization is introduced. 

\section {Tetraquark Group Theory and Pauli restrictions}

We now use group theory to examine the effect of symmetry restrictions from the
Pauli principle on the fragmentation of a $uu\bar u \bar s$ tetraquark with no
orbital angular momentum into a $K^+\pi^o$ state. 

The quark and the antiquark are classified respectively in the sextet and 
antisextet representations of the color-spin $SU(6)$ group, $SU(6)_{cs}$. 
Pseudoscalar mesons are color singlets and spin singlets and are singlets
in $SU(6)_{cs}$. Vector mesons are  color singlets and spin triplets and are 
classified in the $35$ dimensional representation of $SU(6)_{cs}$. Thus the
pseudoscalar-pseudoscalar final states are color singlets and spin singlets and 
singlets in $SU(6)_{cs}$ while the
vector-pseudoscalar states are color singlets and spin triplets and are 
classified in the $35$ dimensional representation of $SU(6)_{cs}$.
The particular simplicity of the pseudoscalar-pseudoscalar final state gives
rise to the unique Pauli restrictions discussed below. These restrictions do not
apply to other states which are not singlets in color spin and $SU(6)_{cs}$.   

States of two quarks are classified in $SU(6)_{cs}$ in either the symmetric 
$6\otimes 6 = 21$ representation or the antisymmetric $6\otimes 6 = 15$
representation. 

The $uu\bar u \bar s$ tetraquark with no orbital angular momentum contains a 
$uu$ pair which is required by the Pauli principle to be in the antisymmetric
15-dimensional representation of $SU(6)_{cs}$. Because the final
two-pseudoscalar meson state is in a spin-zero color-singlet state, the  $\bar
u \bar s$ pair must also be in the 15-dimensional representation. Although no
Pauli principle forbids it from being in the symmetric 21-dimensional
representation, the states in the product $15\otimes 21$ contain no spin-zero
color singlet. Note that the  product $15\otimes 21$ contain a spin-one color
singlet and can be an allowed state for vector-pseudoscalar decays. Thus our
treatment here applies exclusively only to the two-pseudoscalar decay modes.. 

The $uu\bar u \bar s$ tetraquark contains only $u$ and $s$ flavors; its SU(3)
flavor symmetry is conveniently described by using the SU(2) V-spin (us)
subgroup.   Both the $uu$ diquark and the $\bar u \bar s$ antidiquark are
antisymmetric in color-spin. The generalized Pauli principle requires them both
to be in the flavor-symmetric V-spin state with $V=1$. 

The $(V=1,V_z=+1)$ diquark denoted by $\ket{D(1,+1)}$  and the $(V=1,V_z=0)$
antidiquark denoted by $\ket{\bar D(1,0)}$ can be coupled  either 
symmetrically to  a tetraquark denoted by $\ket{T(V=2,V_z=+1)}$ with
total $V$-spin $(V=2,V_z=+1)$ or antisymmetrically to  a tetraquark denoted by 
$\ket{T(V=1,V_z=+1)}$ with
total $V$-spin 
$(V=1,V_z=+1)$.  

\beq{vtetp}
\ket{T(V=2,V_z=+1)} = \ket{D(1,+1;\bar D(1,0)} +
\ket{\bar D(1,+1;D(1,0)}   
\eeq

\beq{vtetm}
\ket{T(V=1,V_z=+1)} = \ket{D(1,+1;\bar D(1,0)} - 
\ket{\bar D(1,+1;D(1,0)}   
\eeq

The tetraquark state must be even under  generalized charge
conjugation to decay into two pseudoscalar mesons. That the state
$\ket{T(V=2,V_z=+1)}$ satisfies this condition and the state 
$\ket{T(V=1,V_z=+1)}$ does not 
can be seen by 
examining the behavior of these states under the $G_{us}$ transformation,
$u\rightarrow \bar s; s\rightarrow \bar u$; i.e. the analog of G-parity using
$V$ spin instead of $I$ spin.

\beq{gvtet}
G_{us}\ket{D(1,+1;\bar D(1,0)} = 
\ket{\bar D(1,+1;D(1,0)}
\eeq 
\beq{gvtet2}
G_{us}\ket{T(V=2,V_z=+1)}  = \ket{T(V=2,V_z=+1)}
\eeq 
\beq{gvtet1}
G_{us}\ket{T(V=1,V_z=+1)}  = - \ket{T(V=1,V_z=+1)}
\eeq 
The state $\ket{T(V=2,V_z=+1)}$ is in the 27-dimensional representation of 
flavor SU(3).

Although the SU(6) color-spin algebra is used in this analysis, there is no
assumption here that the dynamics are invariant under SU(6). The algebra here is
just a short cut for writing down the explicit wave functions and imposing the
restriction of the Pauli principle. The requirement that this $uu\bar u \bar s$ 
tetraquark must be classified in the 27-dimensional representation of flavor 
SU(3) follows from the SU(3) flavor symmetry and the fact that it
has no orbital angular momentum and spin zero and the generalized charge
conjugation with SU(3) that allows the decay into two octet pseudoscalar mesons.     

In contrast the $ud\bar u \bar s$ which is created in the tree diagram for
$B_d$ decay has no such restrictions. In particular it can be in a flavor SU(3)
octet as well as a 27. Its ``diquark-antidiquark" configuration includes the
flavor-SU(3) octet constructed from the   spin-zero color-antitriplet
flavor-antitriplet ``good" diquark found in the  $\Lambda$ baryon and its
conjugate ``good" antidiquark. These ``good diquarks" do not exist in the  
corresponding $uu\bar u \bar s$ tetraquark configuration.

The final $K^+\pi^o$ state is a tetraquark state which is a linear combination
of the $uu\bar u \bar s$ and  $ud\bar d \bar s$ states. This state is seen to
have only a small $V=2$ component.  The $\pi^o$ is (3/4) $V=0$ and only (1/4)
$V=1$. Only the $V=1$ component can couple with the $V=1$ $K^+$ to make $V=2$.
The coupling of a  $(V=1,V_z=+1)$ state to a $(V=1,V_z=0)$ state has an equal
probability of making states with $V=2$ and $V=1$. Thus the probability that
the final  $K^+\pi^o$ state has $V=2$ is (1/8). Only this small $V=2$ component
can contribute to the  fragmentation process which creates the final $K^+\pi^o$
state from the $V=2$ tetraquark in the approximation  where SU(3) is conserved
and therefore also its $V$ spin subgroup. This suppression factor of (1/8) does
not arise in the $B_d$ decays whose final states are expected to be further 
enhanced by their creation from ``good" diquarks in a flavor octet state.  
  
We again see that the Pauli effects produce a drastic symmetry difference
produced by spectator quark flavor on the tree diagrams for $B \rightarrow K
\pi$ decays.

\section {conclusion}

Experiment has shown that the penguin-tree interference contribution in $B^+
\rightarrow K^+\pi^o$ decay is very small and may even vanish. The
corresponding interference contributions to neutral $B\rightarrow K\pi$ decays
have been shown experimentally to be finite. This can explain why CP violation
has been observed in neutral $B \rightarrow K\pi$ decays and not in charged
decays.  A new isospin analysis suggests a possible selection rule justifying
the apparent experimental fact that  the color-favored and color-suppressed
contributions to the amplitude for the $B^+ \rightarrow K^+ \pi^o$ decay seem
to be equal and opposite and cancel. The requirement that identical quarks
appearing in different final state hadrons must satisfy the Pauli principle 
has provided serious constraints in a group-theoretical treatment selecting  a
unique allowed SU(3) flavor final state with V-spin $V=2$.  . This could
provide an additional constraint for analyses of systematics and CP violation
in $B \rightarrow K \pi$ decays. 

\section*{Acknowledgements}

This research was supported in part by the U.S. Department of Energy, Division
of High Energy Physics, Contract W-31-109-ENG-38. It is a pleasure to thank
Michael Gronau, Yuval Grossman, Marek Karliner, Zoltan Ligeti, Yosef Nir,
Jonathan Rosner, J.G. Smith, and  Frank Wuerthwein for discussions and
comments.

%
\catcode`\@=11 
\def\references{
\ifpreprintsty \vskip 10ex
%
\hbox to\hsize{\hss \large \refname \hss }\else
\vskip 24pt \hrule width\hsize \relax \vskip 1.6cm \fi \list
{\@biblabel {\arabic {enumiv}}}
{\labelwidth \WidestRefLabelThusFar \labelsep 4pt \leftmargin \labelwidth
\advance \leftmargin \labelsep \ifdim \baselinestretch pt>1 pt
\parsep 4pt\relax \else \parsep 0pt\relax \fi \itemsep \parsep \usecounter
{enumiv}\let \p@enumiv \@empty \def \theenumiv {\arabic {enumiv}}}
\let \newblock \relax \sloppy
 \clubpenalty 4000\widowpenalty 4000 \sfcode `\.=1000\relax \ifpreprintsty
\else \small \fi}
\catcode`\@=12 

\end{document}